\documentclass[twocolumn]{aastex701}

\usepackage{subcaption}
\usepackage{graphicx}
\usepackage{placeins}
\usepackage{microtype}

\begin{document}

\title{Under Pressure: UV Emission Line Ratios as Barometers of AGN Feedback Mechanisms}

\author[0000-0001-6846-9399]{Elise Fuller}
\affiliation{Department of Astronomy, The University of Michigan, 1109 Geddes Avenue, Ann Arbor, MI, 48109, USA}
\affiliation{Center for Astrophysics and Space Astronomy, Department of Astrophysical and Planetary Sciences, University of Colorado, 389 UCB, Boulder, CO 80309-0389,
USA}
\email{elke1424@colorado.edu}

\author[0000-0001-9487-8583]{Sean D. Johnson}
\affiliation{Department of Astronomy, The University of Michigan, 1109 Geddes Avenue, Ann Arbor, MI, 48109, USA}
\email{seanjoh@umich.edu}

\author[0000-0002-7541-9565]{Jonathan Stern}
\affiliation{School of Physics and Astronomy, Tel Aviv University, 69978 Tel Aviv, Israel}
\email{sternjon@tauex.tau.ac.il}

\author[0000-0001-8813-4182]{Hsiao-Wen Chen}
\affiliation{Department of Astronomy \& Astrophysics, The University of Chicago, 5640 South Ellis Avenue, Chicago, IL 60637, USA}
\affiliation{Kavli Institute for Cosmological Physics, The University of Chicago, 5640 South Ellis Avenue
Chicago, IL, 60637}
\email{hchen@oddjob.uchicago.edu}

\author[0000-0002-8131-6378]{Ena Choi}
\affiliation{Department of Physics, University of Seoul, 163 Seoulsiripdaero, Dongdaemun-gu, Seoul 02504, Republic of Korea}
\email{enachoi@uos.ac.kr}

\author[0000-0002-4900-6628]{Claude-Andr\'e Faucher-Gigu\`ere}
\affiliation{Department of Physics and Astronomy, Northwestern University, 2145 Sheridan Road, Evanston, IL 60208, USA}
\affiliation{Center for Interdisciplinary Exploration and Research in Astrophysics, Northwestern University, 1800 Sherman Ave, Evanston, IL 60208, USA}
\email{cgiguere@northwestern.edu}

\author[0000-0003-2754-9258
]{Massimo Gaspari}
\affiliation{Department of Physics, Informatics and Mathematics, University of Modena and Reggio Emilia, 41125 Modena, Italy}
\email{massimo.gaspari@unimore.it}

\author[0000-0003-4700-663X]{Andy Goulding}
\affiliation{Department of Astrophysical Sciences, Princeton University, 4 Ivy Lane, Princeton, NJ 08544, USA}
\email{goulding@astro.princeton.edu}

\author[0000-0002-5612-3427]{Jenny Greene}
\affiliation{Department of Astrophysical Sciences, Princeton University, 4 Ivy Lane, Princeton, NJ 08544, USA}
\email{jennyg@princeton.edu}

\author{Timothy M. Heckman}
\affiliation{
The William H. Miller III Department of Physics and Astronomy, The Johns Hopkins University, Baltimore, MD 21218, USA}
\email{theckma1@jhu.edu}

\author[0000-0002-0311-2812]{Jennifer~I-Hsiu Li}
\affiliation{Center for AstroPhysical Surveys, National Center for Supercomputing Applications, University of Illinois Urbana-Champaign, Urbana, IL, 61801, USA}
\affiliation{Michigan Institute for Data Science, University of Michigan, Ann Arbor, MI, 48109, USA}
\affiliation{Department of Astronomy, The University of Michigan, 1109 Geddes Avenue, Ann Arbor, MI, 48109, USA}
\email{jennili@umich.edu}

\author[0000-0002-2662-9363]{Zhuoqi Liu}
\affiliation{Department of Astronomy, The University of Michigan, 1109 Geddes Avenue, Ann Arbor, MI, 48109, USA}
\email{zql@umich.edu}

\author[0000-0002-9141-9792]{Nishant Mishra}
\affiliation{Department of Astronomy, The University of Michigan, 1109 Geddes Avenue, Ann Arbor, MI, 48109, USA}
\email{mishran@umich.edu}

\author[0000-0003-1991-370X]{Kristina Nyland}
\affiliation{U.S. Naval Research Laboratory, 4555 Overlook Ave SW, Washington, DC 20375, USA}
\email{kristina.e.nyland.civ@us.navy.mil}

\author[0000-0001-7883-8434]{Kate Rowlands}
\affiliation{William H. Miller III Department of Physics and Astronomy, Johns Hopkins University, Baltimore, MD 21218, USA}
\affiliation{AURA for ESA, Space Telescope Science Institute, 3700 San Martin Dr., Baltimore, MD 21218, USA}
\email{krowlands@stsci.edu}

\author[0000-0002-8459-5413]{Gwen C. Rudie}
\affiliation{The Observatories of the Carnegie Institution for Science, 813 Santa Barbara Street, Pasadena, CA 91101, USA}
\email{gwen@carnegiescience.edu}

\author[0000-0001-9735-7484]{Evan Schneider}
\affiliation{Department of Physics and Astronomy, University of Pittsburgh, Pittsburgh, PA 15260, USA}
\affiliation{Pittsburgh Particle Physics, Astrophysics, and Cosmology Center (PITT PACC), University of Pittsburgh, Pittsburgh, PA 15260, USA}
\email{eschneider@pitt.edu}

\author[0000-0003-2212-6045]{Dominika Wylezalek}
\affiliation{Astronomisches Rechen-Institut, Zentrum für Astronomie der Universität Heidelberg, Münchhofstr. 12-14, 69120 Heidelberg, Germany}
\email{dominika.wylezalek@uni-heidelberg.de}

\author[0000-0001-6100-6869]{Nadia L. Zakamska}
\affiliation{Department of Physics and Astronomy, Bloomberg Center, Johns Hopkins University, 3400 N. Charles Street, Baltimore, MD
21218, USA}
\email{zakamska@jhu.edu}

\begin{abstract}

Feedback from active galactic nuclei (AGN) is widely acknowledged to regulate the growth of massive galaxies, though its driving mechanisms are debated. Prevailing theories suggest that AGN\textcolor{black}{-driven} outflows are driven either by radiation pressure acting directly on the dusty interstellar medium (ISM) or by hot winds entraining cooler ISM gas, but the relative contribution of each mechanism remains uncertain. By combining optical emission line measurements with highly ionized UV emission lines, it is possible to constrain \textcolor{black}{whether the pressure source applied to ionized clouds is primarily radiation or primarily hydrodynamic, and thus constrain the dominant driver.}
This study presents the first multi-object analysis of far-ultraviolet (FUV) spectra from galactic-scale \textcolor{black}{AGN-driven} outflows in obscured quasars, based on Cosmic Origins Spectrograph observations of five low-redshift targets. By comparing narrow-line region UV emission line ratios to theoretical models that vary the importance of the two \textcolor{black}{pressure sources}, we find three out of five targets fall within the radiation pressure-dominated regime. A fourth target exhibits intermediate emission-line ratios that suggest radiation pressure and pressure from a hot wind are both dynamically important. Finally, the lowest-luminosity object in our sample may have a dynamically important hot wind component, but non-detections prevent a clear conclusion in this case. These results suggest radiation pressure dominates circum-nuclear narrow-line region cloud dynamics, but pressure from a hot wind also plays a role in some cases. This is consistent with AGN feedback scenarios mediated by radiation pressure or a short-lived hot wind phase that dissipates after initially accelerating outflows.
\end{abstract}


\section{Introduction}\label{sec:intro}

By incorporating feedback from active galactic nuclei (AGN) as the primary regulator of massive galaxy evolution, many current cosmological simulations successfully reproduce key observable phenomena \citep[see][and references therein]{bessiere_2024}. These include the well-established correlations between the properties of supermassive black holes (SMBH) and their host galaxies \citep{Cattaneo_2009, Kormendy_2013}, the suppression of star formation in massive galaxy halos \citep{Behroozi_2013, Su_2021}, and the exponential break in the galaxy luminosity function \citep{Bower_2006, Weigel_2017}. 

The effectiveness of these models has positioned AGN feedback as a cornerstone of modern theories of massive galaxy evolution \citep[for reviews, see][]{Harrison:2017aa, Morganti:2017aa, Martin_Navarro_2018, Laha:2021aa}. In particular, AGN feedback appears crucial to regulating the sizes of massive galaxies \citep[e.g.][]{Parsotan_2021, Byrne_2024}. However, its driving mechanisms remain enigmatic, especially for radio-quiet quasars which dominate AGN demographics but lack the dramatic radio jets observed driving feedback in rarer radio-loud systems \citep[e.g.][]{hardcastle_2020}. 

\textcolor{black}{For AGN feedback to significantly impact star formation, many models require acceleration of cool and cold interstellar medium (ISM) clouds. During this process, some mechanism must exert substantial pressure, thereby setting the properties of the resulting outflow.} Two primary theories propose \textcolor{black}{the dominant pressure mechanism in radio-quiet quasar outflows as}: (1) direct radiation pressure on cool gas in the \textcolor{black}{ISM} or (2) ISM entrainment by a hot wind originating from the nucleus. Determining which of these is the primary driver is a subject of active research and debate \citep[for a review, see][]{Singha_2023}. 

In radiation pressure-driven outflow models, absorption and scattering of photons in dusty gas surrounding an AGN create pressure, expelling surrounding material as an outflow \citep{Murray_2005, Ishibashi_2018, Arakawa_2022}. In the presence of non-negligible optical depth, radiation pressure also compresses the outflowing gas into a stratified pressure gradient, creating a layered ionization structure with the most highly ionized species in the outermost layers closest to the AGN and lower ionization states in the interior of the clouds \citep{Dopita_2002, Baskin_2013aa, Baskin_2013bb, Stern_2014a,Stern_2014,Bianchi_2019,Netzer_2022}. This depth-dependent gas pressure and density is predicted to produce emission lines across a broad range of ionization states. 

In hot wind-driven AGN outflow models,  quasi-relativistic winds \textcolor{black}{generated within the gravitational zone of influence of the} central SMBH \citep{Kurosawa_2009}, which may originally be driven by radiation pressure, \textcolor{black}{collide with surrounding ISM on larger scales }and produce shocks that propagate in both directions on impact. These shocks produce a hot, volume-filling wind that can contribute to observed outflows by entraining and accelerating the cooler ISM \citep{King_2011, zubovas_2012, Faucher_Gigu_re_2012, Richings_2018, Richings_2018b}. In these models, confining hot gas is expected to \textcolor{black}{exert pressure by} compress\textcolor{black}{ing} the cool, line-emitting gas. Such compression leads to lower \textcolor{black}{and more uniform} expected ionization compared to outflows primarily driven by radiation pressure \citep{Allen_2008, Stern_2016, Richings_2021}. 

Recent surveys largely characterize AGN outflows with strong optical emission lines from low- and intermediate-ionization species, particularly [\ion{O}{3}] \citep[e.g.,][]{Greene_2011, Harrison_2014, Sun_2017, Leung_2019, Meena_2021}. Such lines can identify candidate AGN outflows, but alone, they cannot diagnose their drivers, as they do not measure a wide range of ionization states. 

Predictions of emission line ratios for radiation pressure-dominated outflows are precise, with minimal dependence on model parameters, since in such outflows the incident radiation sets the ionization, thermal, and density structure \citep{Stern_2014a,Stern_2016,Bianchi_2019}. \textcolor{black}{Thus, limits on the ionization parameter, $U$, which sets the emission,} tightly constrain the contribution of hot winds  \citep{Stern_2016}. In particular, UV emission lines such as \ion{O}{6}, \ion{N}{5}, and \ion{C}{4} observable with the Cosmic Origins Spectrograph \citep[COS;][]{Green:2012vw} on the \textit{Hubble Space Telescope} (\textit{HST}) provide powerful diagnostics when combined with optical line observations.

 A pilot study \citep{Somalwar:2020aa} obtained the first spatially resolved UV spectra of a prototypical low-$z$ obscured quasar, J1356$+$1026, to perform these emission line ratio diagnostics. Their {\it HST} COS G140L spectra revealed \ion{O}{6}/\ion{N}{5} line ratios consistent with models of AGN photoionized clouds confined primarily by radiation pressure in both the narrow-line region (NLR) and extended outflow. Building on this pilot study, we present circum-nuclear NLR FUV emission spectra from \textit{HST} COS for a sample of five radio-quiet, low-redshift ($z$) obscured quasars exhibiting galactic-scale outflows in order to enable UV-to-optical emission line diagnostics.

In Section \ref{sec:obs}, we discuss the target selection process, optical SDSS data, and FUV {\it HST} COS observations. Our resulting UV and optical emission line fluxes are discussed in Section \ref{sec:results}, and the models from \cite{Stern_2016} we compare them to are discussed in Section \ref{sec:models}. We examine the observed UV-to-UV and UV-to-optical emission line ratios in Sections \ref{sec:results} and \ref{sec:discussion}. Throughout our analysis, we adopt a flat $\Lambda$ cosmological model with a Hubble constant of $H_0=70$ km/s/Mpc, matter density of $\Omega_{\rm m}=0.3$, and dark energy density of $\Omega_\Lambda=0.7$. 
\renewcommand{\nodata}{}

\section{Observations and Data Reduction} \label{sec:obs}

\subsection{Target Selection}

\begin{deluxetable*}{lrrccrc}
\tablecaption{Summary of {\it HST} COS observations. \label{tab:cos_obs}}
\tablehead{
\colhead{Object} & \colhead{RA} & \colhead{Dec} & \colhead{Redshift} & \colhead{Grating/cenwave} & \colhead{$t_{\rm exp}$} & \colhead{Start} \\
\colhead{} & \colhead{[deg]} & \colhead{[deg]} & \colhead{} & \colhead{} & \colhead{[s]} & \colhead{Date}
}
\startdata
J1356$+$1026 & 209.19208 & 10.43583 & 0.123 & G140L/1105 & 5200 & 2019-06-12  \\ 
J0841$+$0101 & 130.39621 & 1.03231 & 0.111 & G140L/1105 & 480 & 2020-02-26 \\ 
J1000$+$1242 & 150.05475 & 12.70728 & 0.148 & G140L/1105 & 700 & 2020-05-26  \\ 
J1222$-$0007 & 185.57438 & -0.12881 & 0.173 & G140L/1105 & 500 & 2020-05-18  \\
              \nodata & \nodata & \nodata &  \nodata    & G130M/1291 & 500 & 2020-05-18 \\ 
J1255$-$0339 & 193.94933 & -3.65267  & 0.169 & G140L/1105 & 2400 & 2020-07-04\\
              \nodata &  \nodata    & \nodata & \nodata  & G130M/1291 & 2300 & 2020-07-04 
\enddata

\end{deluxetable*}

To investigate the relative importance of radiation pressure and hot winds in AGN feedback (see discussion in Section \ref{sec:intro}), we targeted radio-quiet obscured quasars with known galactic-scale ($\sim$ 10 kpc) outflows for {\it HST} COS observations of highly ionized emission lines (\ion{O}{6}, \ion{N}{5}, and \ion{C}{4}) arising from the circum-nuclear narrow-line region. Comparing these emission lines with photoionization models that vary the relative importance of radiation pressure and hot wind pressure allows us to constrain the density structure within the ionized clouds \citep{Stern_2016}. To ensure these emission features fell within the COS FUV window, we selected quasars with redshifts between $z\approx 0.1$ and $0.17$. We specifically chose obscured quasars because dust in their hosts acts as a natural coronagraph, blocking the bright central nucleus and allowing clearer observations of any extended outflows and the NLR conditions.

We identified targets meeting these criteria from \cite{Greene_2011}, \cite{Harrison_2014}, and \cite{Sun_2017}. These surveys identified candidate extended outflows from Sloan Digital Sky Survey \citep[SDSS;][]{York_2000} Type 2 quasars catalogued by \citep{Reyes_2008, Mullaney_2013}, and characterized them using optical [\ion{O}{3}] emission lines observed with either long-slit or integral field optical spectroscopy. Using radio data from FIRST \citep{FIRST_Becker_1995} and NVSS \citep{NVSS_Condon_1998}, \cite{Greene_2011} and \cite{Harrison_2014} classify their samples as radio-quiet based on their position on the $\nu L_\nu(1.4\,\rm{GHz})-L_{\rm{[O\,III]}}$ plane \citep{Xu_1999,Zakamska_2004}.

Our selection resulted in targets with comparatively high [\ion{O}{3}] luminosities (see the right panel of Figure \ref{fig:BPT_diagram}), consistent with observations that luminous AGN more frequently host outflows \citep{zakamska_2014, Polednikova_2015}. 
All five of our targets show evidence of ongoing or recent merger activity \citep{molyneux_2023,Foord_2020,comerford_2015, pfeifle_2023,Sun_2017,Dutta_2022,Ramos_Almeida_2022}, supporting theories that luminous Type 2 quasars at low- and intermediate redshift are often triggered by gas funneled to the nucleus during galaxy interactions \citep[e.g.][]{Goulding_2018, Pierce_2023}. 

First, we include \textit{J1356$+$1026}, at $z=0.123$, the subject of an \textit{HST} COS pilot study by \cite{Somalwar:2020aa}. It was chosen from \cite{Greene_2011} based on their long-slit observations made with the Low-Dispersion Survey Spectrograph 3 \citep[LDSS3;][]{Allington-Smith_1994} on the Magellan/Clay telescope at
the Las Campanas Observatory in Chile. J1356$+$1026 has the most pronounced extended and kinematically disturbed outflow out of their sample of 15 luminous, obscured, low-$z$ quasars, and they estimate its [\ion{O}{3}] emission radius as $\approx 10$ kpc \citep[also see][]{Greene_2012}. 

We selected two more targets from \cite{Greene_2011}: J1222$-$0007 and J0841$+$0101. \textit{J1222$-$0007}, at $z$=0.173, is a spiral galaxy with an [\ion{O}{3}] emission radius of $\sim 12.6$ kpc and relatively high emission-line width ($>$500 km s$^{-1}$) \citep{Greene_2011}. \textit{J0841$+$0101}, at $z$=0.111, has a reported [\ion{O}{3}] emission radius of $\sim 10$ kpc \citep{Greene_2011}. 

We targeted \textit{J1000$+$1242}, at $z$=0.148, based on the GMOS South IFU observations from \cite{Harrison_2014}. They selected it from the parent catalog of \cite{Mullaney_2013} as part of a representative study of 16 Type 2, low-$z$ obscured quasars. They report an extended [\ion{O}{3}]-emitting region radius of 4.3 $\pm$ 1.8 kpc. 

We targeted \textit{J1255$-$0339}, at $z$=0.169, based on the Magellan long-slit observations in \cite{Sun_2017}. It had the most extended narrow-line region in their sample of 12 nearby luminous obscured AGN, with a $R_{\rm{NLR}}$ (narrow-line region radius) of 33.5 $\pm$ 1.4 kpc.

It is possible for radio-quiet quasars to become radio-loud on $\approx10$ year timescales as new radio jets launch \citep[][]{Nyland_2020}. To account for this possibility, we checked more recent radio observations of our targets from the Very Large Array Sky Survey \citep[VLASS;][]{Lacy_2020}. All five targets are detected in VLASS with flux levels consistent with the previous radio-quiet classifications. However, we note that the objects are not radio-silent, so the presence of low power jets cannot be ruled out.

\subsection{SDSS Optical Data}\label{subsection: optical fluxes}

To ensure a uniform analysis of the sample, which was drawn from a number of AGN catalogs, and to produce the optical emission diagnostics discussed in Section \ref{sec:results}, we performed our own emission line flux measurements of each target's nuclear optical spectra from the SDSS DR17 Data Release \citep{Abdurro_uf_2022}. Specifically, we measured emission in [\ion{N}{2}], H$\alpha$, [\ion{O}{3}], H$\beta$, and H$\gamma$.

Due to blending of the H$\alpha$ and [\ion{N}{2}] lines, we measured optical fluxes by fitting Gaussian profiles to each emission line. We simultaneously fit linear continuum models with Gaussian profiles separately for the H$\alpha$+[\ion{N}{2}], H$\beta$, H$\gamma$+[\ion{O}{3}] $\lambda$4363 \AA, and [\ion{O}{3}] $\lambda\lambda$4960, 5008 \AA\ regions. We modeled both the [\ion{N}{2}] doublet ($\lambda\lambda$6550, 6585 \AA) and [\ion{O}{3}] doublet ($\lambda\lambda$4960, 5008 \AA) with pairs of Gaussians with fixed amplitude ratios of 2.95, based on the ratio of their spontaneous emission rates \citep{Osterbrock_2006}. To find the best-fit model parameters for each region, we applied a $\chi^2$ minimization routine with the \texttt{lmfit} module \citep{newville_2015_11813}. Due to degeneracies between line fluxes, widths, and redshifts of individual Gaussian components, we evaluated the total flux posteriors of each emission feature using Markov Chain Monte Carlo (MCMC) methods from the \texttt{emcee} module \citep{2013PASP..125..306F} with walker start positions set around the best fit from the $\chi^2$ minimization routine.

We constrain the H$\alpha$, H$\beta$, and H$\gamma$ Balmer lines to share component parameters since they originate from the same ionized hydrogen gas undergoing recombination, and thus experience the same broadening mechanisms. Specifically, their Gaussians shared redshifts and widths (i.e., we modeled broad components of H$\alpha$, H$\beta$, and H$\gamma$ with the same line width). This choice was also motivated by the significant H$\alpha$+[\ion{N}{2}] blending, as fitting to the unblended H$\beta$ and H$\gamma$ lines helped inform the H$\alpha$ models. We used two shared-width components -- one broad and one narrow -- to model H$\alpha$, H$\beta$, and H$\gamma$ for J0841$+$0101 and J1000$+$1242. J1356$+$1026, J1222$-$0007, and J1255$-$0339 exhibited significant blueshifted emission components indicative of a wind, fit by an ``additional" component with a FWHM slightly smaller than that of the narrow components. Modeling the J1356$+$1026 wing accurately required two ``additional" components. We note that the total line fluxes considered in our analysis do not depend significantly on the detailed breakdown of the emission lines into Gaussian components.

We fit the [\ion{N}{2}] $\lambda\lambda$6550, 6585 \AA\ doublet together with H$\alpha$ in order to capture uncertainty due to blending of [\ion{N}{2}] and H$\alpha$ in the model posteriors, though we modeled each emission line individually. The [\ion{N}{2}] doublet was effectively modeled with a single Gaussian for J0841$+$0101, J1356$+$1026, and J1222$-$0007, and with two Gaussian components --- one broad and one narrow --- for J1000$+$1242 and  J1255$-$0339.

We fit the [\ion{O}{3}] $\lambda\lambda$4960, 5008 \AA\ doublet separately from the other observed emission lines, with its own continuum and component structure. One broad and one narrow Gaussian effectively modeled this doublet for J1000$+$1242 and J1356$+$1026. An additional component was necessary for J1255$-$0339, J0841$+$0101, and J1222$-$0007. Again, J1255$-$0339, J1356$+$1026, and J1222$-$0007 exhibit blueshifted wings in their [\ion{O}{3}] emission, modeled with an ``additional" component for J1222$-$0007 and J1255$-$0339 but captured by the broad component for J1356$+$1026.

\begin{figure*}[htbp]
\includegraphics[width=0.9\textwidth]{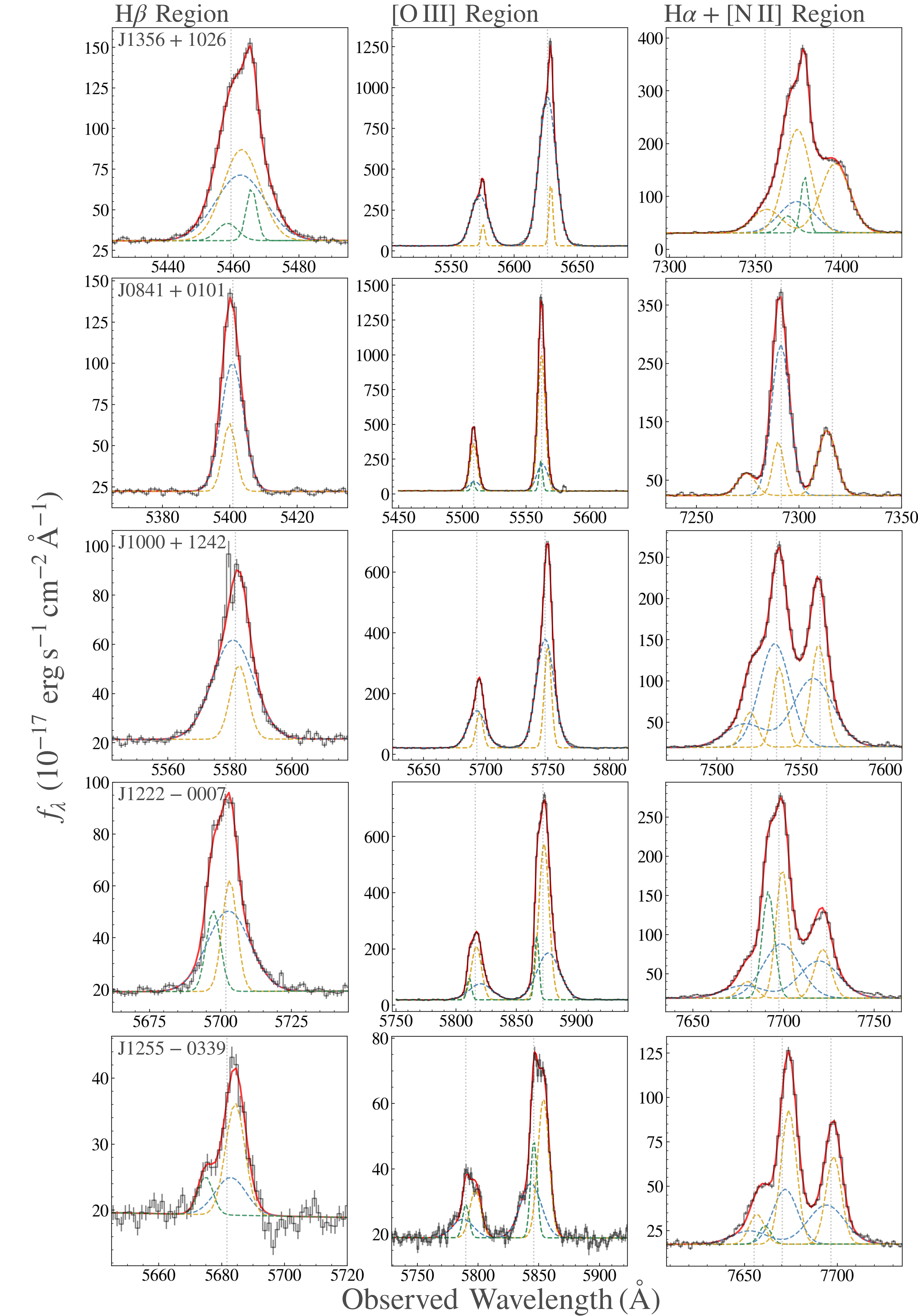}
    \caption{Gaussian fits to SDSS optical emission lines for our five target AGN. ``Narrow" components are in gold, ``broad" components in blue, and ``additional" components in green. Emission line regions span similar ranges between objects of  $\approx$2000 km/s (H$\beta$ region), $\approx$5000 km/s ([\ion{O}{3}] region), and $\approx$2500 km/s (H$\alpha$+[\ion{N}{2}] region) on either side of the central emission line (or the midpoint between both, in the case of [\ion{O}{3}]). For more on these fits, see Section \ref{subsection: optical fluxes}.}
    \label{fig:allsdssfits}

\end{figure*}
\begin{figure*}[htbp]
\includegraphics[width=0.95\textwidth]{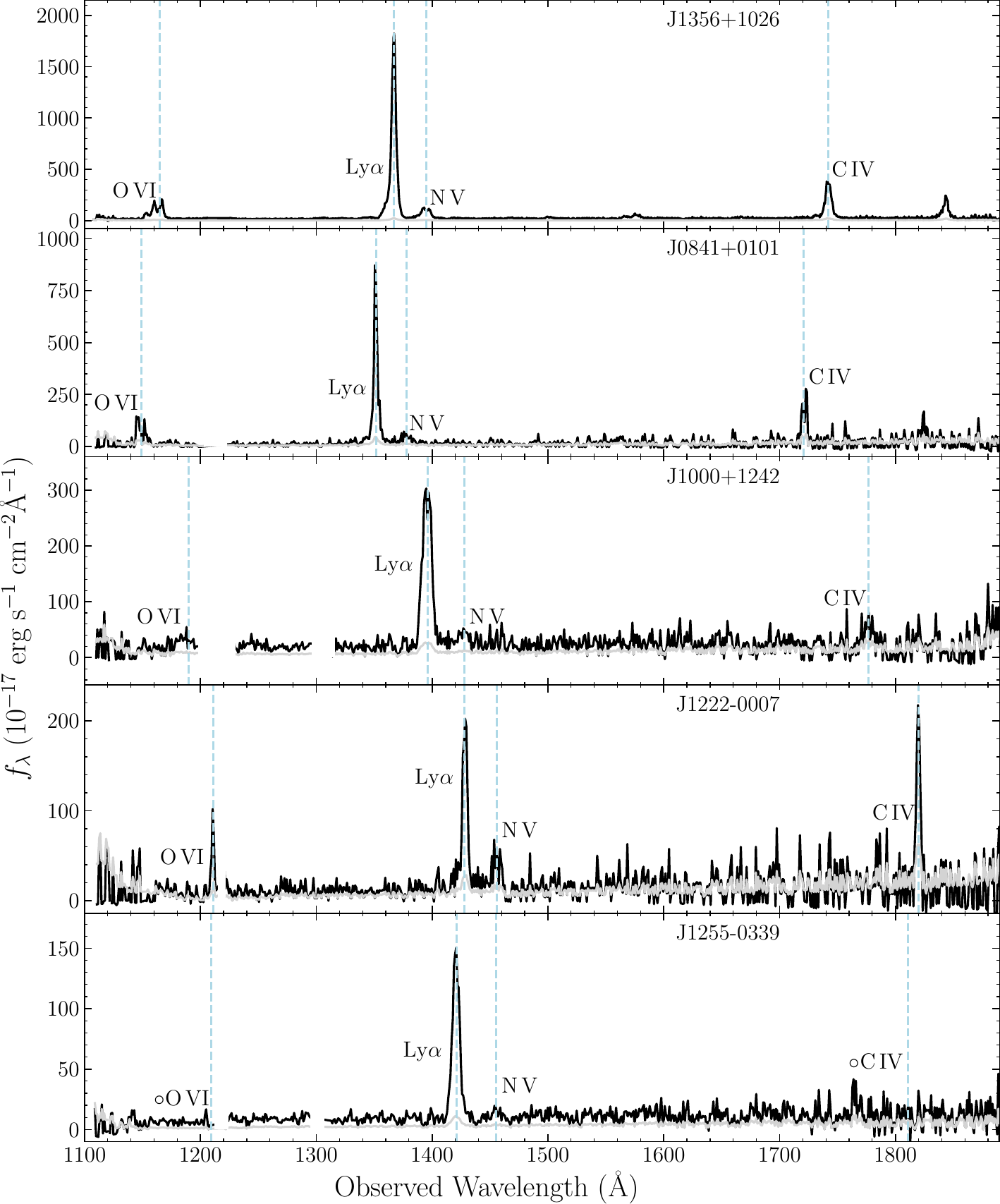}
    \caption{COS UV nuclear spectra for our five AGN targets, with flux in black and its uncertainty in gray. Dashed light blue lines mark the expected location of the labeled spectral line given the target's redshift. For observed doublets (\ion{O}{6} $\lambda\lambda$1031, 1037 $\textrm{\AA}$, \ion{N}{5} $\lambda\lambda$1238, 1242 $\textrm{\AA}$, and \ion{C}{4} $\lambda\lambda$1548, 1550 $\textrm{\AA}$) the blue line is located at the midpoint of the expected location of each emission line. The two reported non-detections, \ion{O}{6} and \ion{C}{4} for J1255$-$0339, are denoted on their respective labels with open circles. To aid the visual inspection of emission lines, we have masked out the geocoronal Ly$\alpha$ and O\,I lines, leaving gaps at approximately 1215 and 1304 \AA, where applicable}.
    \label{fig:alluvspectra}

\end{figure*}

The SDSS optical spectra and best-fitting  Gaussian models are shown in Figure \ref{fig:allsdssfits}. The optical line fluxes and uncertainties reported in Table \ref{tab:cos_fluxes} are based on the total line flux posteriors marginalized over other parameters. 

\subsection{{\it HST} COS Observations and Data Reduction} \label{subsection: uv data}

To cover the wavelength range needed for UV emission-line diagnostics, we obtained far-ultraviolet (FUV) nuclear spectra for our five targets from {\it HST} COS. The COS observations of J1356$+$1026 come from the pilot program described in Section \ref{sec:intro} \citep[PID 15280]{Somalwar:2020aa}. COS observations of the other four obscured quasars came from a follow-up survey (PID 15935). All observations used for this analysis are summarized in Table \ref{tab:cos_obs}. 

The new {\it HST} COS observations were taken  primarily with the G140L 1105 \AA\, grating, but also supplemented with G130M 1291 \AA\, in some cases. Both gratings were imaged with all four FP-POS positions. G140L has a relatively low resolution (resolving power ranging from 1,500--4,000) and was chosen for its wide wavelength range. Our G140L spectra cover wavelengths between $\sim$1100--1900 $\textrm{\AA}$. We also obtained observations with the G130M grating since its higher resolution (resolving power ranging from 12,000--16,000) enables better separation of the O\,VI line from geocoronal H\,I Ly$\alpha$. We spliced the G130M spectra into the G140L spectra in the region surrounding geocoronal \ion{H}{1} Ly$\alpha$ emission for two objects with nearby \ion{O}{6} emission (J1222$-$0007 and J1255$-$0339). 

We retrieved the 1D spectrum associated with each exposure reduced using the CALCOS pipeline \citep[Version 3.4.3;][]{Kaiser_2008} from the MAST archive. We then combined the individual exposures into final coadded spectra after masking out pixels with data quality flags. In particular, we included pixels with no anomalies and those flagged as being in the detector wire shadow, as reduction flat-fielding is sufficient to correct any artifacts from this at our observed $S/N$ \citep{Soderblom2022COS}. We co-added spectra onto a rebinned wavelength grid with a pixel scale of 1 $\textrm{\AA}$ for all five of the targets. This new pixel scale was courser than either G140L or G130M, and flux calibrations were consistent between both  gratings, so we did not need to perform resolution or flux matching on the spliced G140L + G130M spectra. Flux and counts were conserved throughout the co-adding process.

In order to correct for the effects of Milky Way dust extinction, we used the extinction curve from \cite{Fitzpatrick_1999} implemented in the \texttt{Extinction} package \citep{barbary_2021} and $A_V$ values estimated in the direction of each obscured quasar from the extinction map in \cite{Schlafly_2011}. The resulting coadded and extinction-corrected spectra are shown in Figure \ref{fig:alluvspectra}. Within these spectra, we observe emission in the \ion{O}{6} $\lambda\lambda$1031, 1037 $\textrm{\AA}$ doublet; the \ion{N}{5} $\lambda\lambda$1238, 1242  $\textrm{\AA}$ doublet; and the \ion{C}{4} $\lambda\lambda$1548, 1550 $\textrm{\AA}$ doublet. The process we used to measure these emission line fluxes is described in Section \ref{subsection: uv fluxes}, and the results are reported in Section \ref{sec:results}.

\begin{deluxetable*}{lccccccc}
\tablecaption{UV and optical nuclear emission line fluxes, counts, observed centroids, signal-to-noise ratios, full width at half maximums, and line-of-sight velocities relative to systemic redshift. \label{tab:cos_fluxes}}

\tablenum{2}
\tablehead{\colhead{} & \colhead{Line} & \colhead{Flux} & \colhead{Counts} & \colhead{Centroid} & \colhead{SNR} & \colhead{FWHM} & \colhead{$\Delta V$} \\ 
\colhead{} & \colhead{} & \colhead{(10$^{-15}$ erg cm$^{-2}$ s$^{-1}$)} & \colhead{} & \colhead{(\AA)}  & \colhead{} & \colhead{(km/s)}  & \colhead{(km/s)} }
\startdata
J1356$+$1026   & H I Ly$\alpha$ & 96.3{\hspace{0.2mm}}$\pm0.5$ & 47661 & 1366 & 238 & 960 & 177 \\
{}   & \ion{N}{5} 1238/1242\tablenotemark{a} & 9.7{\hspace{0.2mm}}$\pm0.2$ & 6394 & 1391 & 82 & 2080 & 7 \\
{}   & \ion{O}{6} 1031/1037\tablenotemark{a} & 16.8{\hspace{0.2mm}}$\pm0.3$ & 4089 & 1162 & 70 & 2508 & 11 \\
{}   & \ion{C}{4} 1548/1550\tablenotemark{a} & 25.5{\hspace{0.2mm}}$\pm0.6$ & 4017 & 1742 & 56 & 884 & 304  \\
{}   & [\ion{O}{3}] 5008 & 124.3{\hspace{0.2mm}}$\pm0.4$ & --- & 5626 & 311 & 663 & 154 \\
{}   & [\ion{N}{2}] 6584 & 20.2{\hspace{0.2mm}}$\pm0.3$ & \nodata & 7394 & 67 & 473 & 10 \\
{}   & H$\beta$ & 19.5{\hspace{0.2mm}}$\pm0.2$ & \nodata & 5462 & 98 & 905 & 138 \\
{}   & H$\alpha$ & 64.2{\hspace{0.2mm}}$\pm0.5$ & \nodata & 7374 & 128 & 882 & 138 \\ \hline 
J0841$+$0101   & H I Ly$\alpha$ & 37.8{\hspace{0.2mm}}$\pm1.5$ & 796 & 1351 & 29 & 644 & 256 \\
{}   & \ion{N}{5} 1238/1242\tablenotemark{a} & 4.9{\hspace{0.2mm}}$\pm0.5$ & 123 & 1376 & 11 & 2545 & 68 \\
{}   & \ion{O}{6} 1031/1037\tablenotemark{a} & 9.2{\hspace{0.2mm}}$\pm1.0$ & 106 & 1149 & 11 & 2169 & 5 \\
{}   & \ion{C}{4} 1548/1550\tablenotemark{a} & 11.6{\hspace{0.2mm}}$\pm1.5$ & 71 & 1721 & 7 & 966 & 105 \\
{}   & [\ion{O}{3}] 5008 & 74.7{\hspace{0.2mm}}$\pm0.3$ & --- & 5560 & 249 & 349 & 5 \\
{}   & [\ion{N}{2}] 6584 & 9.3{\hspace{0.2mm}}$\pm0.1$ & \nodata & 7311 & 93 & 433  & 1\\
{}   & H$\beta$ & 10.1{\hspace{0.2mm}}$\pm0.1$ & \nodata & 5400 & 101 & 411& 74 \\
{}   & H$\alpha$ & 36.2{\hspace{0.2mm}}$\pm0.2$ & \nodata & 7290 & 181 & 400 & 75\\ \hline 
J1000$+$1242   & H I Ly$\alpha$ & 28.3{\hspace{0.2mm}}$\pm1.0$ & 957 & 1397 & 32 & 2131 & 240\\
{}   & \ion{N}{5} 1238/1242\tablenotemark{a} & 3.0{\hspace{0.2mm}}$\pm0.3$ & 169 & 1426 & 13 & 2480 & 229\\
{}   & \ion{O}{6} 1031/1037\tablenotemark{a} & 5.2{\hspace{0.2mm}}$\pm0.4$ & 220 & 1187 & 16 & 4139 & -200\\
{}   & \ion{C}{4} 1548/1550\tablenotemark{a} & 9.0{\hspace{0.2mm}}$\pm1.1$ & 93 & 1778 & 7 & 3523 & -296\\
{}   & [\ion{O}{3}] 5008 & 66.0{\hspace{0.2mm}}$\pm0.3$ & --- & 5747 & 220 & 884 & -74\\
{}   & [\ion{N}{2}] 6584 & 38.7{\hspace{0.2mm}}$\pm0.3$ & \nodata & 7557 & 129 & 598  & -94\\
{}   & H$\beta$ & 10.4{\hspace{0.2mm}}$\pm0.1$ & \nodata & 5582 & 104 & 718 & 11\\
{}   & H$\alpha$ & 38.5{\hspace{0.2mm}}$\pm0.4$ & \nodata & 7536 & 96 & 962 & 11\\ \hline 
J1222-0007   & H I Ly$\alpha$ & 11.7{\hspace{0.2mm}}$\pm0.9$ & 244 & 1427 & 15 & 1046 & 335 \\
{}   & \ion{N}{5} 1238/1242\tablenotemark{a} & 4.0{\hspace{0.2mm}}$\pm0.5$ & 85 & 1455 & 9 & 2020 & -136 \\
{}   & \ion{O}{6} 1031/1037\tablenotemark{a} & 3.7{\hspace{0.2mm}}$\pm0.8$ & 120 & 1214 & 9 & 2078\tablenotemark{c} & 44 \\
{}   & \ion{C}{4} 1548/1550\tablenotemark{a} & 9.2{\hspace{0.2mm}}$\pm1.6$ & 46 & 1819 & 5 & 661 & 304 \\
{}   & [\ion{O}{3}] 5008 & 82.0{\hspace{0.2mm}}$\pm0.3$ & --- & 5871 & 273 & 511 & -93 \\
{}   & [\ion{N}{2}] 6584 & 13.6{\hspace{0.2mm}}$\pm0.2$ & \nodata & 7719 & 68 & 264 & -111 \\
{}    & H$\beta$ & 11.2{\hspace{0.2mm}}$\pm0.1$ & \nodata & 5701 & 112 & 792 & -53 \\
{}   & H$\alpha$ & 45.8{\hspace{0.2mm}}$\pm0.4$ & \nodata & 7696 & 115 & 691 & -53 \\ \hline 
J1255-0339   & H I Ly$\alpha$ & 10.9{\hspace{0.2mm}}$\pm0.3$ & 1342 & 1420 & 36 & 1531& -110 \\
{}   & \ion{N}{5} 1238/1242\tablenotemark{a} & 0.9{\hspace{0.2mm}}$\pm0.4$ & 307 & 1454 & 16 & 3510 & 713 \\
{} & \ion{O}{6} 1031/1037\tablenotemark{a}\tablenotemark{b}   & $<0.9$ & $<$178 & 1209  & ---     & ---  & ---  \\ 
{} & \ion{C}{4} 1548/1550\tablenotemark{a}\tablenotemark{b}  & $<37$ & $<$171 & 1811 & ---   & ---   & ---  \\ 
{}   & [\ion{O}{3}] 5008 & 7.8{\hspace{0.2mm}}$\pm0.1$ & ---  & 5847 & 78 & 325 & $-252$ \\
{}   & [\ion{N}{2}] 6584 & 7.9{\hspace{0.2mm}}$\pm0.2$ & \nodata & 7694 & 40& 356  & -20 \\
{}   & H$\beta$ & 3.0{\hspace{0.2mm}}$\pm0.1$ & \nodata & 5681 & 112 & 411 & $-51$ \\
{}   & H$\alpha$ & 14.5{\hspace{0.2mm}}$\pm0.3$ & \nodata & 7669 & 48 & 400 & $-51$ \\
\enddata
\tablenotetext{a}{Reported values for flux, error, centroid, FWHM, and velocity shift are for the full doublet, not individual lines.}
\tablenotetext{b}{Upper-limit flux estimated as the three sigma limit from Gaussian fits or direct integration as described in Section \ref{subsection: uv fluxes}.}
\tablenotetext{c}{FWHM of a double-Gaussian fit to both \ion{O}{6} lines, assuming the 1037 emission line flux is 0.75 that of the 1031 line (see Section \ref{subsection: uv fluxes} for more on this object's \ion{O}{6} doublet).}
\vspace{-10pt}
\end{deluxetable*}

\subsection{UV Flux Measurements}\label{subsection: uv fluxes}

Since the Ly$\alpha$, \ion{O}{6}, \ion{N}{5}, and \ion{C}{4} UV emission lines are generally not blended with strong lines from other species, we measured their fluxes by applying direct Simpson integration to each individual emission line after subtracting a linear continuum fitted to nearby line-free spectral regions. We do not treat individual doublet lines separately; i.e. the reported doublet emission flux values are the sum of both lines. 

We estimated statistical flux errors using Poisson counting statistics. For verification, we calculated the standard deviation of the mean flux per pixel from surrounding line-free continuum areas for each line and confirmed these were consistent with our Poisson estimates. To account for systematic uncertainties in defining integration and continuum region boundaries, we implemented a Monte Carlo error estimation, which generally contributed an additional $\sim$5-10\% uncertainty to our reported errors on UV flux. 
Our reported count values are the sum of gross counts present within the line integration region. We note that while our flux measurements for J1356$+$1026
are consistent with those from \cite{Somalwar:2020aa}, our count estimates differ due to updates to the CALCOS pipeline and differences in bad pixel masking.

The close proximity of the observed  \ion{O}{6} $\lambda\lambda$1031, 1037  $\textrm{\AA}$ doublet to geocoronal Ly$\alpha$ resulted in potential blending for J1222$-$0007 and J1000$+$1242, despite the use of higher-resolution G130M data. To determine the extent of blending for each, we compared the Full Width at Half Maximum (FWHM) of the other measured doublets (\ion{N}{5} $\lambda\lambda$1238, 1242  $\textrm{\AA}$ and \ion{C}{4} $\lambda\lambda$1548, 1550 $\textrm{\AA}$) to the expected doublet separation. For J1000$+$1242, we found the \ion{O}{6} doublet had a FWHM comparable to or greater than the other measured doublets, and concluded that it is distinctive from geocoronal Ly$\alpha$. For J1222$-$0007, we found that the emission blueward of geocoronal Ly$\alpha$ at the expected location of \ion{O}{6} had the smallest FWHM despite having the largest expected doublet separation, indicating that it was the 1031 emission line, and the 1037 emission line was blended with geocoronal Ly$\alpha$. To verify this and examine the extent of the blending, we fit two Gaussians --- one to each emission line in the doublet, with a shared emission line width --- to the \ion{N}{5} $\lambda\lambda$1238, 1242  $\textrm{\AA}$ and \ion{C}{4} $\lambda\lambda$1548, 1550 $\textrm{\AA}$ doublets, then fit Gaussians with widths fixed at the \ion{N}{5} velocity dispersion  to the \ion{O}{6} $\lambda\lambda$1031, 1037  $\textrm{\AA}$ region. The resulting fit captured the 1031 line and placed the 1037 line within geocoronal Ly$\alpha$, as expected.

Given these results, we estimated the total \ion{O}{6} flux for J1222$-$0007 by considering the theoretical expectation for ratios between the \ion{O}{6} 1031 and 1037 emission lines. In the optically thin limit, the 1031 emission line is expected to be approximately twice as strong as the 1037 line, but in the optically thick limit there is an approximately 1:1 ratio \citep{Draine_2011}. Blending of the 1037 line with geocoronal \ion{H}{1} Ly$\alpha$ prevents measurements of the line ratio, but the theoretical minimum and maximum enables us to place a robust constraint on the total O\,VI flux from the 1031 line alone. In particular, we report the measured total O\,VI flux as 1.75 times the 1031 flux measurement with an associated systematic uncertainty of 0.25 times the 1031 flux measurement.

We report upper limits on the possible flux for the two emission lines that did not meet our detection threshold of $\geq 3\sigma$, \ion{C}{4} and \ion{O}{6} in J1255$+$0339's spectrum. We estimated an upper-limit flux for \ion{C}{4} by calculating the average flux per count multiplied by the gross counts within 23 $\textrm{\AA}$ --- the width of this object's Ly$\alpha$ emission line --- of the expected line position. For \ion{O}{6}, this method was unreliable due to contamination from geocoronal Ly$\alpha$, so we instead adopted a $3\sigma$ upper limit from a double Gaussian fit with redshift fixed to the SDSS value and velocity dispersion fixed to that of the AGN's Ly$\alpha$. 

Table \ref{tab:cos_fluxes} summarizes the optical and UV line fluxes and signal-to-noise ratios for our targets, along with counts for the UV lines only. The measurements and line ratios are further shown and discussed in Sections \ref{sec:results} and \ref{sec:discussion}.

\begin{figure*}
   \centering
    \includegraphics[width=\linewidth]{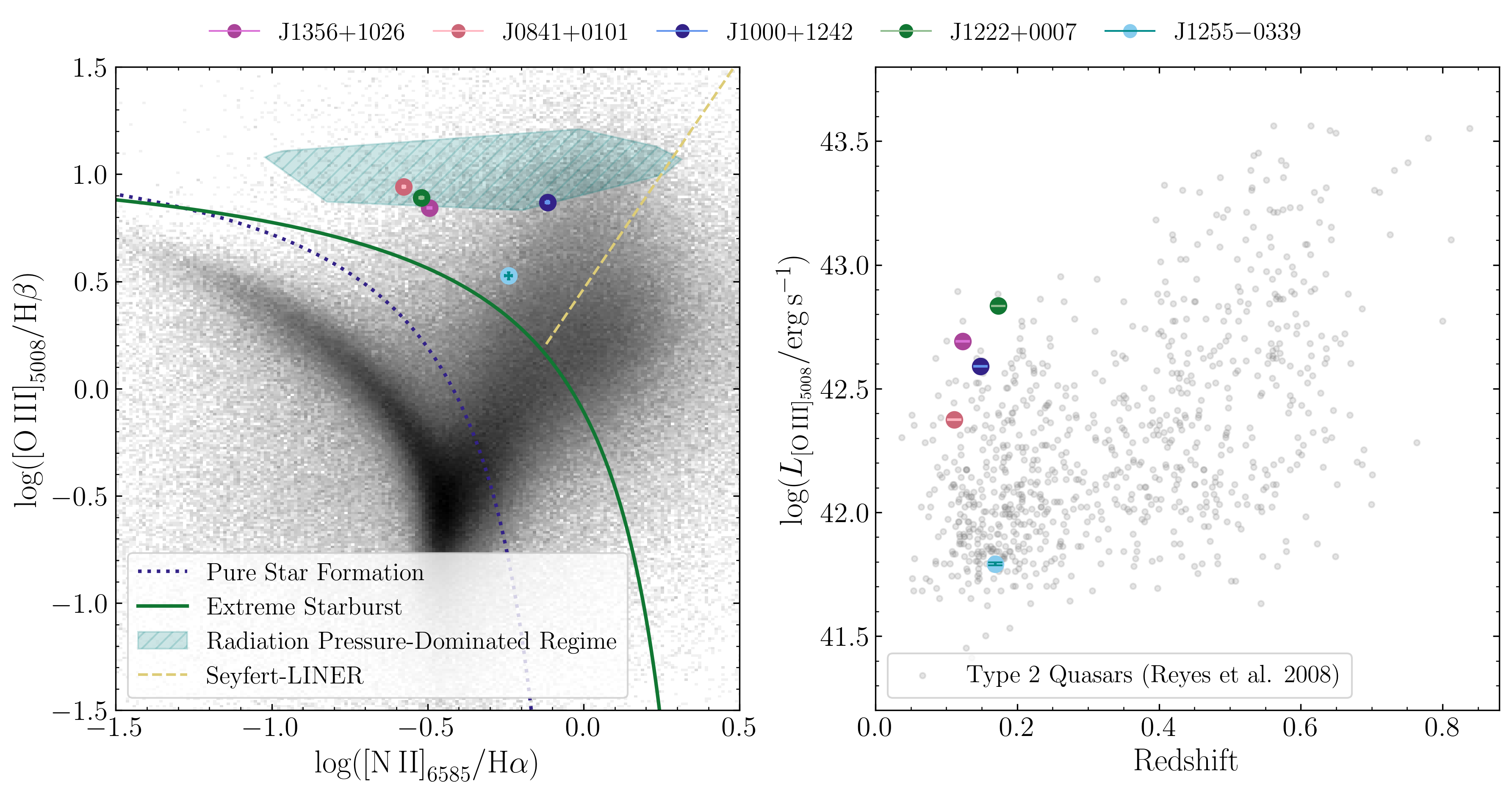}
    \caption{\textit{Left panel:} BPT diagram confirming identification of objects as AGN, with target obscured quasars as colored points and SDSS galaxies as small grey points. The solid green and dotted blue lines are the extreme starburst and pure star formation classification lines from \cite{Kewley:2006wa}. The Seyfert-LINER line from \cite{Kauffmann_2003} is a dashed yellow-orange line. The range predicted by \textcolor{black}{radiation pressure-dominated} models ($P_{\rm rad}\gg P_{\rm hot}$) discussed in Section \ref{sec:models} for the NLR is shown as a hatched teal region. Targets are shown as multi-colored points, while error estimates are smaller than the datapoints. \textit{Right panel:} [\ion{O}{3}] luminosities measured from SDSS spectra versus redshift for our five targets plotted against those of the Type 2 Quasars catalog presented in \cite{Reyes_2008}. The 68\% confidence interval errors, plotted in similar colors atop the data, are smaller than the data points. }
    
    \label{fig:BPT_diagram}
\end{figure*}

\subsection{Aperture Effects}\label{subsection: aperture}

Both the SDSS optical data and HST + COS UV data are calibrated to account for flux losses from point sources. The UV data was taken with the 2.5'' diameter \textit{HST} COS aperture with high angular resolution, and modulated by a space-based point spread function (PSF).
The SDSS spectra were collected with fibers under ground-based seeing conditions with fiber sizes ranging from $2''$ to $3''$.
In particular, the SDSS spectra for four out of five targets (J1356$+$1026, J1000$+$1242, J1222$-$0007, and J1255$-$0339) were taken with the original SDSS spectrograph (3'' diameter aperture), and the optical spectrum for J0841$+$0101 was taken with BOSS (2'' aperture diameter). The majority of our targets' narrow-line region emission is expected to arise from within the central $\lesssim$ 100 pc, a prediction based on the efficiency of [\ion{O}{3}] emission in gas with radii down to $\approx$ 20 pc $L_{46}^{1/2}$ \citep{Stern_2014a} and validated by HST observations of NGC 1068, the only Type 2 quasar close enough to resolve such scales with \textit{HST}, which show [\ion{O}{3}] emission dominated by radii $\lesssim$ 80 pc \citep{Revalski_2021}. Thus, for our aperture scales of 2'' -- 3'', the majority of the narrow-line region flux likely arises from unresolved spatial scales near the nucleus, corresponding to  $\approx$  0.05'' at $z$=0.111 and $\approx$ 0.03'' at $z$=0.173. However, the narrow-line region can extend to much larger, galactic scales, and may contribute non-negligible additional flux to the SDSS fiber spectra and COS UV spectra. Thus, we analyze here how much additional flux from non-point sources may be included within each aperture, and how this could affect flux ratios between the UV and optical datasets.

To quantify how much our extended sources deviate from the assumed point source calibration, we obtained data from the Multi-Unit Spectroscopic Explorer \citep[MUSE;][]{Bacon_2010} for J1356$+$1026 (Program ID 0103.B-0071; PI C. Harrison), J1000$+$1242 (Program IDs 0103.B-0071 and 0104.B-047; PIs C. Harrison and G. Venturi), and J1222$+$0007 (Program ID 0103.B-0071; PI C. Harrison). For each MUSE dataset, we compared the curve-of-growth of NLR emission lines to that of a nearby point source reference star at different apertures. We examined the [\ion{O}{3}] emission line for J1356$+$1026 (FWHM of 1.25'') and J1000$+$1242 (FWHM of 1.32'') within a 10'' diameter aperture --- the maximum region possible surrounding the J1000$+$1242 reference star, which was on the edge of the image. Since [\ion{O}{3}] emission was absent for J1222$+$0007, we examined its H$\alpha$+[\ion{N}{2}] region (FWHM of 0.96'') within a 6'' diameter aperture, the maximum possible region surrounding the reference star without contamination from a neighboring source. The MUSE observations indicate that the optical flux in the SDSS fibers is dominated by the unresolved narrow-line region, with non-point-source flux contribution at the level of $\sim$10 \% within the 3'' aperture, $\sim$7\% within the 2.5'' aperture, and $\sim$5 \% within the 2.0'' aperture. We therefore conclude that aperture effects are a sub-dominant contribution to the UV-to-optical line ratio uncertainties, which are dominated by systematics in ground-based spectrophotometry. We note that these estimates are based entirely on optical emission lines. Ideally, a similar experiment would be performed on UV emission-line images, but this would require new synthetic narrow-band imaging of UV lines \citep[e.g.][]{Hayes_2016, Ha_2025}, which are not available for our targets.

\section{AGN photoionization models}\label{sec:models}

\textcolor{black}{Here, we describe how our photoionization models account for the pressure sources discussed in Section \ref{sec:intro}, following the approach of \cite{Stern_2016}. We first consider an AGN-driven outflowing cloud of gas at $T\sim10^4\,$K that is embedded within a hot medium and exposed to radiation from the quasar. 
The surrounding hot medium compresses the cloud, exerting gas or ram pressure and establishing the boundary condition at its illuminated face. Deeper into the cloud, radiation pressure from gas and dust opacity further compresses the gas as more and more photons from the quasar are absorbed. Modeling these outflowing clouds requires considering compression from both pressure sources.} 

\textcolor{black}{The expected thermal pressure structure ($P_{\rm gas}$) of the gas cloud depends on which of these pressure sources dominate. In a hot wind-dominated cloud where $P_{\rm hot}\gg P_{\rm rad}\equiv L/(4\pi r^2c)$, we expect a roughly uniform pressure structure throughout the cloud ($P_{\rm{gas}}\sim P_{\rm{hot}}$), since any additional compression due to radiation is small relative to compression by the surrounding hot medium. In contrast, in the radiation pressure-dominated case ($P_{\rm rad}\gg P_{\rm hot}$), pressure from photon absorption builds up within the ionized region and increases with depth into the cloud from the illuminated face. This leads to layered ``slabs" of gas with $P_{\rm{gas}}\sim P_{\rm{hot}}$ in the slab closest to the AGN and $P_{\rm{gas}}\sim P_{\rm{rad}}$ at the \ion{H}{1} front \citep[see][]{Stern_2016,Somalwar:2020aa}.}

\textcolor{black}{To include the effects of radiation pressure in our hydrostatic {\sc CLOUDY} \citep{Ferland_2017} models, we solve for local photoionization and thermal equilibrium in each slab. In addition, these models increase the pressure between consecutive gas slabs according to the following equation from \cite{Pellegrini07}:
\begin{equation}
    \frac{dP_{\rm gas}}{dx} = \frac{\int F_\nu n_{\rm H}\sigma_\nu d\nu}{c} ~,
\end{equation}
where $x$ is the depth into the cloud measured from the illuminated surface, $F_\nu$ is the flux density at frequency $\nu$, $n_{\rm H}$ is the gas density, and $\sigma_\nu$ is the total opacity per hydrogen atom from all absorption and scattering processes in the gas and dust grains. Dust grains are assumed to be fully coupled to the gas via Coloumb forces. Equivalent calculations have also been implemented in the MAPPINGS code by \cite{Dopita_2002} and \cite{Groves_2004}.}

\textcolor{black}{Insight into the relation between our models and observations can be deduced from the well-known relation \citep[e.g.][]{Krolik_1999} between $P_{\rm rad}/P_{\rm gas}$ and the dimensionless ionization parameter, $U$, which quantifies the degree of gas cloud ionization and can thus be constrained by ratios between emission line strengths:}
\begin{equation}
    U\equiv\frac{\int_{\nu_0}\frac{F_\nu}{h\nu}d\nu}{n_{\rm H}c}
\end{equation}
\textcolor{black}{The ionization parameter is then the ratio between ionizing photon flux and the gas density, where $\nu_0$ is the Hydrogen ionization frequency.
At a layer in the \ion{H}{2} region of the cloud with temperature $T$ we thus get:}
\begin{equation}
    \frac{P_{\textrm{rad}}}{P_{\textrm{gas}}}=\frac{\beta\int_{\nu_0}F_\nu/cd\nu}{2n_HkT}=30U\frac{\beta}{2}\left(\frac{T}{10^4\textrm{K}}\right)^{-1}\frac{\langle h\nu\rangle_i}{36\rm{eV}}
\label{eq:pressure_U}
\end{equation}
\textcolor{black}{Here, $\beta$ is a factor of order unity that accounts for absorption of non-ionizing photons by dust and $\langle h\nu\rangle_i$ is the mean energy of an ionizing photon.} 

\textcolor{black}{As stated above, when hot gas is the dominant pressure source, $P_{\rm gas}\approx P_{\rm hot}\gg P_{\rm rad}$ throughout the ionized layer, and equation~(\ref{eq:pressure_U}) yields $U\ll0.03$. In the other limit where $P_{\rm rad}\gg P_{\rm hot}$ --- the radiation pressure-dominated case --- equation~(\ref{eq:pressure_U}) implies that $U$ decreases with increasing depth into the cloud from the ionized face as a result of the pressure and hence density build-up, reaching $U\approx0.03$ at the \ion{H}{1} ionization front where $P_{\rm rad}\approx P_{\rm gas}$. Thus, optical lines which originate from gas with $U<0.1$ are largely emitted from higher-pressure regions near the ionization front, and high ionization UV lines which originate from gas with $U\gtrsim0.1$ are emitted largely from regions closer to the exposed surface of the cloud where gas pressure is lower.}


\textcolor{black}{
As mentioned above, our CLOUDY calculations assume that the ionized gas layer of the emission line cloud achieves hydrostatic equilibrium with both pressure sources. Assuming hydrostatic equilibrium in the emission-line gas is plausible since the sound-crossing time of an ionized layer with thickness $N_{\rm HII}/n_{\rm H}$ is $\sim30\, n_3^{-1}\,{\rm kyr}$ for gas densities of $n_{\rm H}=10^3n_3\,{\rm cm}^{-3}$ and typical columns of $N_{\rm HII}=10^{21}\,{\rm cm}^{2}$ are substantially shorter than outflow evolution and cloud destruction times of 1 Myr or more \citep[see][]{Dopita_2002,Namekata_2014}. A quasi-hydrostatic pressure profile where the force on the ionized layer is roughly balanced by ram pressure from gas beyond the ionization front is thus expected throughout most of the clouds lifetime, even in gas which forms an outflow.  The ram pressure on the back side can be either due to the neutral shielded part of the cloud or due to the ambient medium through which the cloud is moving. }

\textcolor{black}{We note also that in both the radiation pressure-dominated and hot wind pressure-dominated models, the line-emitting clouds are expected to be smaller than $\sim$100 pc \citep{Stern_2014}, and hence all emitting slabs in a cloud will fit within the COS aperture. This allows for comparison of our measurements to the model flux ratios, which are integrated across the entire cloud. }

\textcolor{black}{Previous theoretical investigations of such photoionized hydrostatic AGN clouds demonstrated that the integrated UV emission-line ratios can be used to infer $P_{\rm rad}/P_{\rm hot}$. For comparison, we also consider commonly used `constant density' models, which do not solve the hydrostatic equation and instead assume the photoionized gas has some uniform density. We note that constant density models are very similar to the hydrostatic models in the $P_{\textrm{hot}}$-dominated limit.}

\begin{figure*}
    \centering
    \includegraphics[width=1\linewidth]{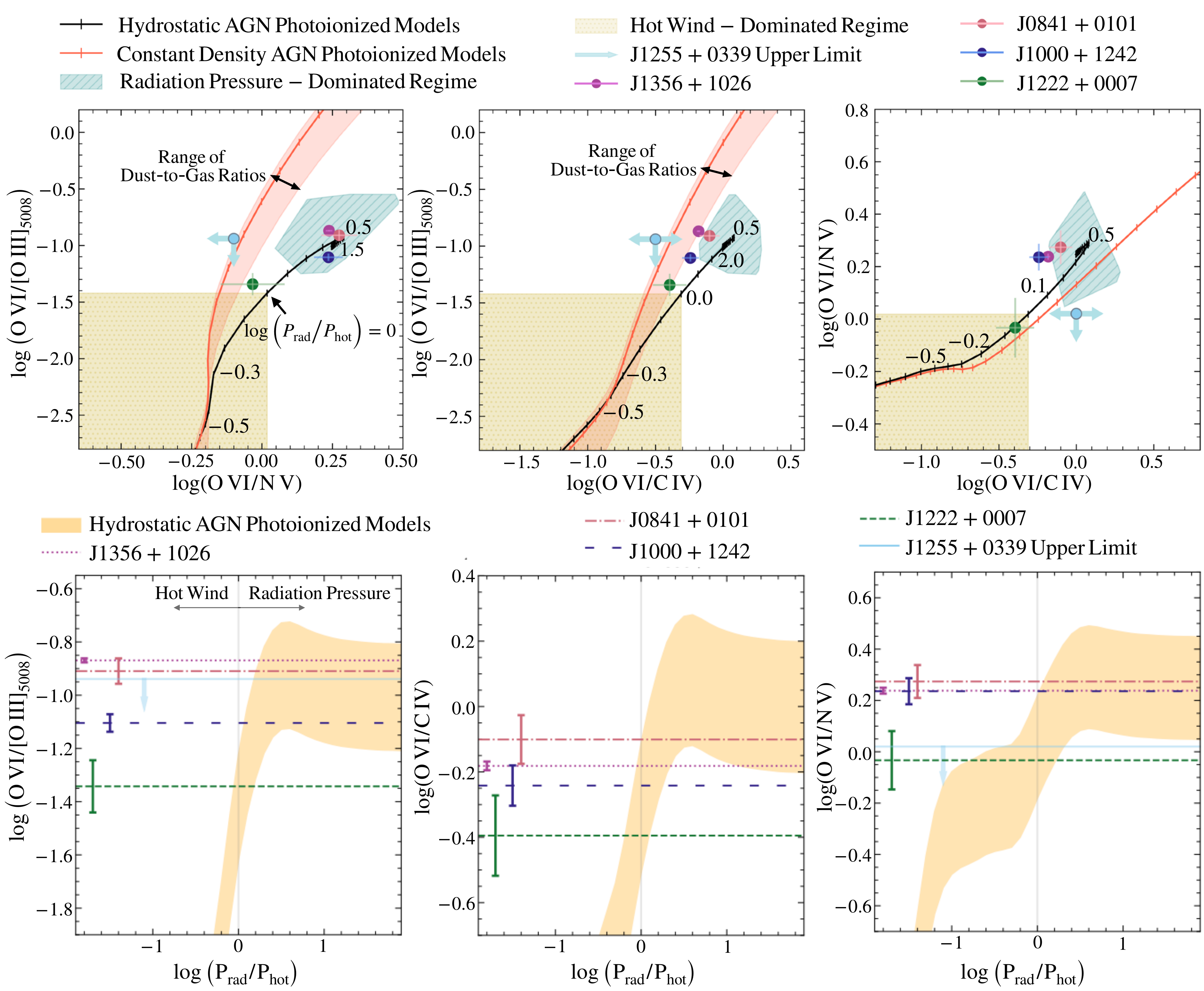}
    \caption{ {\it{Top}}: Observed UV and optical diagnostic line ratios and errors for our five obscured quasars, shown as colored points ({\it{Left}}: \ion{O}{6}/[\ion{O}{3}]$_{5008}$ vs. \ion{O}{6}/\ion{N}{5}, {\it{Middle}}: \ion{O}{6}/[\ion{O}{3}]$_{5008}$ vs. \ion{O}{6}/\ion{C}{4}, {\it{Right}}: \ion{O}{6}/\ion{N}{5} vs. \ion{O}{6}/\ion{C}{4}). Upper-limit ratios from the J1255$-$0339 non-detections of \ion{O}{6} and \ion{C}{4} are shown as light blue points with arrows. Hydrostatic models of AGN photoionized gas varying $P_{\rm rad}/P_{\rm hot}$ from \cite{Stern_2016} with $Z/{\rm Z}_\odot=2{\rm Z}_\odot$,  $\alpha_{\rm ion}=-1.6$ and $r\approx1\,{\rm kpc}$ are shown in solid black, with ticks marking steps of $P_{\rm rad}/P_{\rm hot}=0.1\,{\rm dex}$. The teal, shaded regions plot the predicted ranges for $P_{\rm rad} \gg P_{\rm hot}$ models with $1 < Z/{\rm Z}_\odot < 4$, $-1.8 < \alpha_{\textrm{ion}} < -1.4$, and $100\,{\rm pc} < r < 10\,{\rm kpc}$.
    A rough estimate of the hot wind-dominated regime ($P_{\rm hot} > P_{\rm rad}$) is shown in light yellow. Constant density models are shown in red and at high density are equivalent to hydrostatic models with $P_{\rm hot}\gg P_{\rm rad}$
    The constant density model values within the same range of dust-to-gas ratios as the hydrostatic models are plotted as light red bands for log(\ion{O}{6}/[\ion{O}{3}]) only, as it is impacted most by the extinction effects discussed in Section \ref{subsec:dust}. \textcolor{black}{Three out of five} objects are close to the radiation pressure-dominated limit in all three panels, with two exceptions (J1222$-$0007 and J1255$-$0339). {\it{Bottom}:} Predicted line ratios versus the relative importance of radiation pressure and hot gas pressure for the hydrostatic photoionization models from \cite{Stern_2016}, shown with the UV diagnostic line ratios and errors for the sample of five obscured quasars. Each object is plotted with a different color to aid visual inspection. Their errors are shown as bars on the left side of their respective plots at arbitrary x-axis positions chosen to aid visibility. The legend in the upper left panel applies to all three plots. The J1255$-$0339 upper-limit \ion{O}{6} and \ion{C}{4} flux estimates are shown with a dashed light blue line and arrows in the left and bottom right diagnostic plots, but omitted from the top right plot since both \ion{O}{6} and \ion{C}{4} are non-detections for this object. In all three plots, a light grey vertical line indicates the $\log(P_{\rm rad}/P_{\rm hot})=0$ models.}
    \label{fig:line_ratios}
\end{figure*}

\textcolor{black}{Model parameters in the Cloudy calculation are set to the following fiducial values, and are common between the hydrostatic and constant density models.} We assume a cloud that is optically thick to hydrogen ionizing photons at a distance of $r=1\,{\rm kpc}$ from the AGN, an AGN luminosity of $L=10^{46}\,{\rm erg}\,{\rm s}^{-1}$, and an ionizing spectral slope $\alpha_{\rm ion}=-1.6$ ($L_\nu\propto\nu^{\alpha_{\rm ion}}$ at wavelengths 6.2--1100\AA).  The spectral shape in non-ionizing frequencies follows \cite{Laor_1993}, though its effect on the predicted lines is minor. 
We use the standard Milky-Way ISM grain model implemented in {\sc cloudy} with a graphite and silicate mixture and an MRN size distribution \citep{Mathis_1977}, and metal abundances from \cite{groves_2006} including depletion onto dust grains and non-linear scaling of Nitrogen with metal mass $Z$ (non-depleted abundances are consistent with \citealt{Asplund_2009}). While there is a similar nonlinear relation between C/O and O/H abundance that our models do not account for, the trend is weaker than the relative abundance scaling for nitrogen relation \citep[e.g.][]{Gustafsson_1999, Akerman_2004, Nissen_2014}. Our fiducial models assume twice the metal mass fraction and dust-to-gas mass ratio, $\mathcal{D}$, relative to Milky-Way ISM values, as observed in the centers of massive galaxies which are characteristic of AGN hosts \citep{Kewley:2006wa,groves_2006}. To assess the sensitivity of our results to these `nuisance' parameters, we run a grid of models with $-1.8<\alpha_{\rm ion}<-1.4$, $100\,{\rm pc}<r<10\,{\rm kpc}$, and $Z,\mathcal{D}$ between one and four times the Milky-Way ISM value. The models produced by these parameter ranges are shown as teal regions in Figure \ref{fig:line_ratios}. We demonstrate \textcolor{black}{in Section \ref{sec:discussion}} that the choice of these parameters has far less of an effect on the predicted line ratios than the assumed $P_{\rm rad}/P_{\rm hot}$ in the hydrostatic models or the assumed ionization parameter $U\propto L/(n_{\rm H}r^2)$ in the constant density models.

\textcolor{black}{Other models for AGN outflows, such as that of \cite{Murray_1995} and \cite{Kurosawa_2009b}, explain line emission as arising from volume-filling gas which is optically thin to UV photons. Optically thin models typically do not account for the compressing effect of radiation pressure, since this effect is a result of radiation-absorbing gas being pushed against shielded gas as mentioned above. 
Volume-filling models for narrow emission line gas are disfavored by the small filling factors inferred from photoionization calculations \citep[e.g.,][]{Osterbrock_2006}. The optically thin assumption is disfavored by significant line-emission from neutral gas tracers (e.g. \ion{O}{1} 6300\AA) detected in all five targets,} \textcolor{black}{and by the lack of detected [Ca II] $\lambda$7291 emission in all five targets, which suggests depletion onto dust grains that enhance the optical depth by orders of magnitude \citep{NetzerLaor03,Ferguson97,Stern_2014a}. Specifically, upper limits on [Ca II] $\lambda$7291 in all objects are a factor of at least $5$ below predictions of  dustless photoionization models that reproduce the observed [NII]/H$\beta$, disfavoring a dust-less scenario for the NLR.} 

\textcolor{black}{A dusty NLR also disfavors a scenario where the cool, outflowing NLR clouds arose from gas cooling out of a previously accelerated hot wind, since dust grains are expected to undergo rapid destruction due to sputtering in such a wind.}

\section{Results}\label{sec:results}

To confirm the AGN nature of our objects and contextualize them within the general AGN population, we produced a BPT \citep[][]{Baldwin_1981} diagram (Figure \ref{fig:BPT_diagram}, left panel) with diagnostic boundaries from \cite{Kewley:2006wa}, separating AGN from star-forming galaxies on a plot of [N II]/H$\alpha$ vs [\ion{O}{3}]$_{5008}$/H$\beta$ for both the targets and the SDSS galaxy catalog \citep{Abdurro_uf_2022,Brinchmann_2004,Kauffmann_2003,Salim_2007,Tremonti_2004}.  The BPT diagram confirms that all five targets fall within the expected range for Type 2 AGN. Also included on this plot are ratios given by the radiation pressure-dominated hydrostatic photoionized models of the narrow-line region. Four out of five targets fall within the \textcolor{black}{radiation pressure-dominated} predictions, but one object (J1255$-$0339) has a lower $\log$([\ion{O}{3}]$_{5008}$/H$\beta)$ ratio than the \textcolor{black}{radiation pressure-dominated} expectation. In the right panel of Figure \ref{fig:BPT_diagram}, we plot the [\ion{O}{3}] ${5008}$ \AA\, line luminosities versus redshifts for our sample alongside the obscured quasar catalog in \cite{Reyes_2008}. 

The UV spectra exhibit notable variations in line strengths across the sample. To characterize each object's spectrum, we compare the spectral line profiles of our targets to those of J1356$+$1026 (top panel of Figure \ref{fig:alluvspectra}), which was previously examined in \cite{Somalwar:2020aa}. The line ratios in J0841$+$0101 and J1000$+$1242 closely resemble those of J1356$+$1026, although J1000$+$1242 displays noticeably broader spectral features. J1222$-$0007 exhibits a relatively high \ion{C}{4} to Ly$\alpha$ flux ratio compared to the other targets. J1255$-$0339, in contrast, shows generally weak UV emission, with non-detections of \ion{C}{4} and \ion{O}{6}. We provide upper-limit flux estimates for these undetected emission lines based on the expected location and line widths inferred from detected UV lines (see Section \ref{subsection: uv fluxes} for details on this estimation). 

The SDSS optical spectra of J1356$+$1026, J1222$-$0007, and J1255$-$0339 exhibited blueshifted wings on the H$\beta$, [\ion{O}{3}], and H$\alpha$ emission lines. In the UV, the \ion{O}{6}, Ly$\alpha$, \ion{N}{5}, and perhaps \ion{C}{4} emission lines for J1356$+$1026 all exhibit blueshifted wings. J1222$-$0007 appears to have a small, broad blueshifted component on  Ly$\alpha$, but exhibits no discernible wings on the other observed UV emission lines. The Ly$\alpha$ and \ion{N}{5} lines for J1255$-$0339 do not include any clear blueshifted component. We do not differentiate between components in our reported UV flux measurements, as they are the result of direct integration of the entire visible line. Future analysis of the relationship between blueshifted emission wings observed in the optical compared to those in more highly ionized UV lines may be insightful, particularly if the optical spectra include [\ion{Ne}{5}] emission, which is not covered by SDSS for some of our targets.


\section{Discussion}\label{sec:discussion}

\subsection{Diagnostic Line Ratios}\label{subsec:ratios}
In Figure \ref{fig:line_ratios}, we compare our measured optical and UV emission line ratios between \ion{O}{6}, [\ion{O}{3}]$_{5008}$, \ion{N}{5}, and \ion{C}{4} to the hydrostatic and constant density models described in Section \ref{sec:models} and infer constraints on $P_{{\rm rad}}/P_{{\rm hot}}$.
We note two important points to keep in mind when evaluating our diagnostics. (1) Complete models of AGN outflows may be driven by a combination of radiation pressure and hot winds, with different mechanisms dominating in different phases of evolution or spatial scales. The emission line diagnostics performed here thus probe the  ``instantaneous" relative pressure contributions of these mechanisms. (2) Here, we deduce the source of observed pressure, which is not necessarily equivalent to determining the driving mechanisms of the outflow. It is likely that both mechanisms played roles in the initial outflow driving, and the current dominant pressure source is what is being probed here.

In the top left panel of Figure \ref{fig:line_ratios}, the constant density models are largely ruled out by measured values of log(\ion{O}{6}/\ion{N}{5}) for four targets (J1222$-$0007, J1356$+$1026, J0841$+$0101, and J1000$+$1242). Limits on these line ratios for the fifth target (J1255$-$0339) are consistent with either the constant density or hydrostatic models. The \ion{O}{6}/\ion{N}{5} estimate for J1255$-$0339 falls where $\log({{P}_{\rm{rad}}/P_{\rm{hot}}}) < 0$, implying that hot wind pressure dominates if this object's line-emitting clouds are in hydrostatic equilibrium. J1222$-$0007 exhibits \ion{O}{6}/[\ion{O}{3}]$_{5008}$ and \ion{O}{6}/\ion{N}{5} line ratios $\approx 0.2$ dex lower than the radiation pressure-dominated models predict, and best matches models with contributions from hot wind pressure and radiation pressure that are comparable to one another, both hydrostatic and constant density. For the remaining three targets (J1356$+$1026, J0841$+$0101, and J1000$+$1242), the observed \ion{O}{6}/[\ion{O}{3}]$_{5008}$ and \ion{O}{6}/\ion{N}{5} line ratios fall closest to the radiation pressure-dominated hydrostatic model predictions $\left(P_{{\rm hot}}\ll P_{{\rm rad}}\right)$. 

In the top middle panel of Figure \ref{fig:line_ratios}, the observed \ion{O}{6}/[\ion{O}{3}]$_{5008}$ ratios for J1356$+$1026, J0841$+$0101, and J1000$+$1242 fall within the radiation pressure-dominated range. The \ion{O}{6}/\ion{C}{4} ratio for J0841$+$0101 also falls within the radiation pressure-driven model regime, but J1356$+$1026 and J1000$+$1242 exhibit \ion{O}{6}/\ion{C}{4} ratios $\lesssim 0.1$ dex lower than the radiation pressure confinement models predict. The remaining target (J1222$-$0007) exhibits \ion{O}{6}/[\ion{O}{3}] and \ion{O}{6}/\ion{C}{4} ratios lower than the \textcolor{black}{radiation pressure-dominated} models predict by $\approx$ 0.2 dex. Either the hydrostatic or constant density models could predict all flux ratios shown in this plot except J0841$+$0101, which may be best described by the hydrostatic models.

As the aperture corrections resulting from the analysis in Section \ref{subsection: aperture} would be negligible, we do not apply them to our UV-to-optical flux ratios --- rather, we include a comparison with only UV lines in the right panel of Figure \ref{fig:line_ratios}, which largely reflects the results of the other panels. The observed \ion{O}{6}/\ion{C}{4} and \ion{O}{6}/\ion{N}{5} line ratios for J0841$+$0101, J1000$+$1242, and J1356$+$1026 fall within the radiation pressure-dominated regime. Here, the hydrostatic models best describe these objects. J1222$-$0007 falls $\gtrsim2$ dex lower than the radiation pressure-driven model regime, indicating the possible presence of a hot wind component, although its diagnosis remains ambiguous. It could be described well by either the constant density or hydrostatic photoionized models. The upper limit on \ion{O}{6}/\ion{N}{5} for J1255$-$0339 shows that although either confining mechanism could be dynamically important, there could be contributions from a hot wind phase.

To aid the interpretation of these diagnostics, we characterize each ratio individually in the bottom panel of Figure \ref{fig:line_ratios}. For J1222$-$0007, the uncertainty ranges for \ion{O}{6}/[\ion{O}{3}]$_{5008}$ and \ion{O}{6}/\ion{N}{5} consistently fall within the hydrostatic model estimates, but include both the hot wind and radiation pressure-dominated regimes, so neither possibility can be ruled out. Individual line ratios for J1000$+$1242 fall within the $\log{P_{\rm{rad}}/P_{\rm{hot}}} \gtrsim 0$ regime for all line ratios, although it slightly favors the radiation pressure-dominated models. The J1255$+$0339 upper limits again place it within either regime. J1356$+$1026 and J0841$+$0101 consistently fall within the range predicted by the hydrostatic radiation pressure-dominated models.

Overall, three out of five targets (J1356$+$1026, J0841$+$0101, and J1000$+$1242) exhibit emission-line ratios consistent with hydrostatic \textcolor{black}{radiation pressure-dominated} model predictions. This finding echoes results from the pilot study \citep{Somalwar:2020aa}, which also demonstrated that the J1356$+$1026 line ratios fall near the radiation pressure-dominated limit. These three exhibit relatively high [\ion{O}{3}] luminosities compared to the general population of Type 2 Quasars, but constitute the middle luminosities of our sample. The line ratios for J1222$-$0007 fall around $\log(P_{\textrm{rad}}/P_{\textrm{hot}})=0$, with both the radiation pressure-dominated regime and hot wind-dominated regime within the range of uncertainty. However, the results slightly favor the \textcolor{black}{radiation pressure-dominated} model, with ratios skewing towards the $\left(P_{\textrm{hot}}< P_{\textrm{rad}}\right)$ range. We note that J1222$-$0007 has the highest luminosity and highest redshift ($z$=0.173) of our sample.
 
J1255$+$0339 is the least luminous and second-highest redshift ($z$=0.169) quasar in our sample and exhibits the weakest UV emission. Concluding what regime the J1255$+$0339 line ratios fall within is not possible with current data due to the non-detections of \ion{C}{4} and \ion{O}{6}. This object may be best described by the constant density models, although it still could fall within the hydrostatic predictions. Furthermore, the optical diagnostic ratios in the BPT diagram (left panel of Figure \ref{fig:BPT_diagram}) show J1255$+$0339 as the only target in our sample to fall outside the radiation pressure-dominated model expectations. 

\begin{figure*}
    \centering
    \includegraphics[width=1\linewidth]{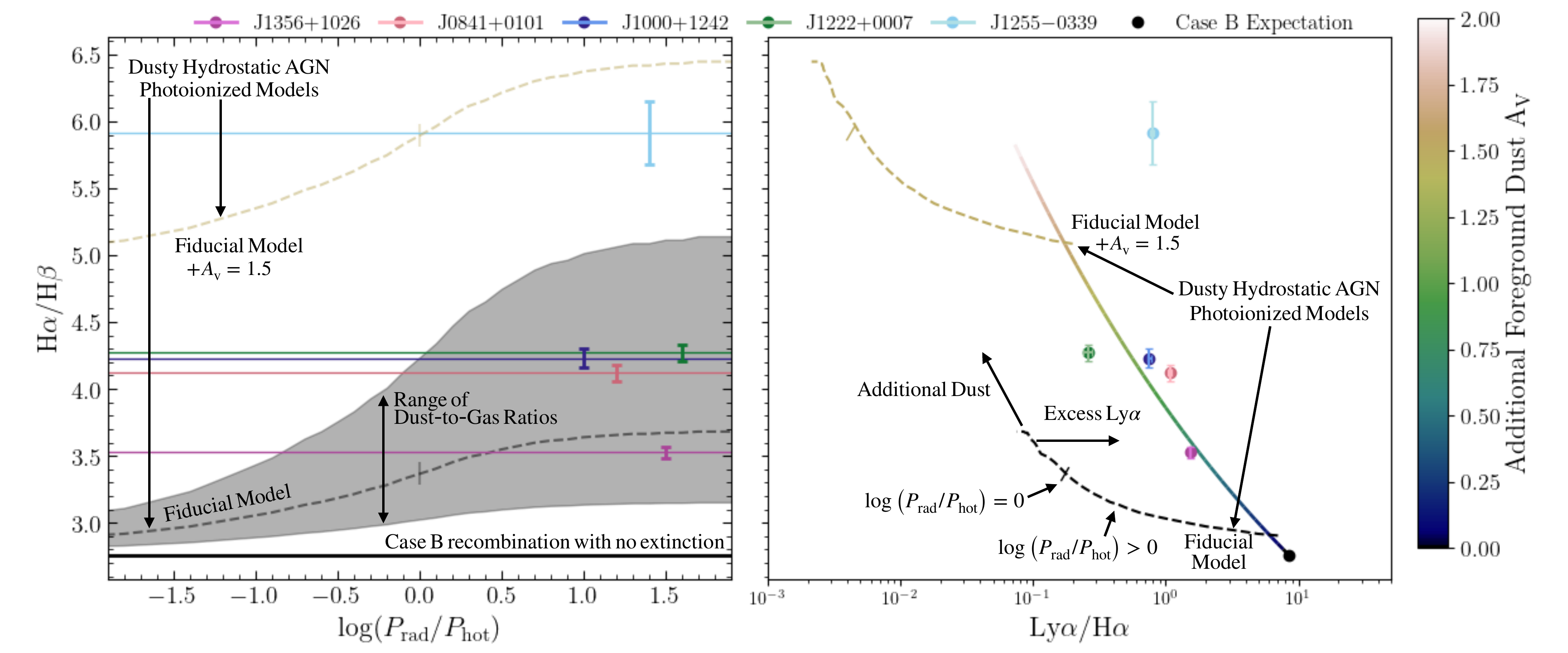}
    \caption{Measured Balmer decrement (H$\alpha$/H$\beta$) ratios versus model $\log (P_{\rm rad}/P_{\rm hot})$(left) and Ly$\alpha$/H$\alpha$ (right), with fluxes corrected for stellar absorption. For reference, both plots include the expected Case B nebula recombination values, as a solid black line in the left panel and a black point in the right panel. To indicate how the line ratio evolves if Case B recombination emission  is subject to an external screen of dust characterized by $A_{\rm V}$, the left panel includes a black arrow and the right panel shows a colored line with $A_{\rm V}$ indicated by the color bar. Also included on both plots are dashed black lines representing the $P_{\rm rad}/P_{\rm hot}$ hydrostatic photoionized AGN models which include dust internal to the NLR clouds, with a shaded grey region representing ratios given by the range of dust-to-gas ratios in the $P_{\rm{rad}}\gg P_{\rm{hot}}$ models described in the caption of Figure \ref{fig:line_ratios}. Also shown as a dotted gold line is the fiducial model with an additional external dust screen of $A_{\rm V}$=1.5, which best matches the ratios of J1255$-$0339.}
    \label{fig:lya_ha}
\end{figure*}

\subsection{Hydrogen Emission Line Ratios \& Dust Extinction}\label{subsec:dust}
Since radiative transfer effects due to dust are a key component of the dusty hydrostatic models, we aim to validate them by checking ratios between available Balmer series and Ly$\alpha$ lines. The optical Balmer emission measurements may be suppressed by Balmer stellar absorption that the models do not account for, preventing accurate comparison. We correct for this by fitting each object's optical continuum spectrum observed by SDSS to a model double power law galaxy stellar population using \texttt{BAGPIPES} \citep{Carnall_2018} after masking out observed emission lines and fitting for dust attenuation using a Calzetti dust law with metallicity fixed at 1 $Z_\odot$, then adjusting the emission flux upward based on the model expectation for stellar Balmer absorption. The resulting corrections are relatively small (typically adjusting fluxes by $5-10$\%). We apply these stellar absorption corrections only to measurements presented in both panels of Figure \ref{fig:lya_ha}, maintaining uncorrected values in Figure \ref{fig:BPT_diagram} and Table \ref{tab:cos_fluxes} for consistency with standard practices in the literature. 

In Figure \ref{fig:lya_ha}, we display the observed H$\alpha$/H$\beta$ and Ly$\alpha$/H$\alpha$ ratios of the targeted AGN and compare these to the expected Case B nebula recombination values assuming a temperature of $T=20,000$\,K from \cite{Osterbrock_2006} and as calculated with \texttt{PyNeb} \citep{Luridiana_2014}. For comparison, we also plot the theoretical evolution of these ratios with increasing dust attenuation (represented by an arrow in the left panel and a colored line spanning $A_V$ values from 0 to 2 in the right panel), alongside our hydrostatic, photoionized model line ratios with additional dust extinction applied. 

On the left panel of Figure \ref{fig:lya_ha}, we include a dust-diagnostic plot of H$\alpha$/H$\beta$ versus $\log(P_{\rm rad}/P_{\rm hot})$ for the dusty hydrostatic models compared to the Case B recombination values, the fiducial hydrostatic model with varying $P_{\rm rad}/P_{\rm hot}$, and the $P_{\rm{rad}}\gg P_{\rm{hot}}$ models covering a range of dust-to-gas ratios described in the caption of Figure \ref{fig:line_ratios}. Four of our five targets (J1356$+$1026, J0841$+$0101, J1000$+$1242, and J1222$+$0007) exhibit H$\alpha$/H$\beta$ ratios consistent with the extinction levels expected from the radiation pressure dominated hydrostatic models. The Balmer emission line ratios of J1255$+$0339 fall outside of this range however, requiring an additional extinction of A$_\textrm{v}\gtrsim 1.5$ above the levels of the fiducial model.

While the H$\alpha$/H$\beta$ ratios diagnose the presence of dust, the Ly$\alpha$/H$\alpha$ ratio --- as a comparison between a resonant and non-resonant line --- shows the impact of additional radiative transfer effects \citep{Scarlata_2009}. The right panel of Figure \ref{fig:lya_ha} shows H$\alpha$/H$\beta$ versus Ly$\alpha$/H$\alpha$ observed for our targets in comparison to expectations for recombination radiation in the presence of dust as well as for the hydrostatic models. The Ly$\alpha$/H$\alpha$ ratios of three out of five targets (J0841$+$0101, J1356$+$1026, and J1000$+$1242) fall within $\approx$ 3$\sigma$ of the Case B expectation with additional foreground dust at levels implied by  H$\alpha$/H$\beta$. However, they exhibit $\gtrsim 2$ dex of excess Ly$\alpha$ when compared to the hydrostatic models with non-negligible radiation pressure, which is indicated by the UV metal line ratios for most of the targets. It is possible for Ly$\alpha$ photons to take random walks and escape from the central region, causing lower-than-predicted measurements of Ly$\alpha$, but this contrasts with the observed excess Ly$\alpha$ relative to the \textcolor{black}{radiation pressure-dominated} models. This excess may be the result of more complicated radiative transfer, requiring more realistic geometries to simulate, or could result from additional sources of Ly$\alpha$ photons such as star formation or scattered light from the broad-line region. The significance of radiative transfer effects for the Ly$\alpha$ emission is supported by the observed Ly$\alpha$ line widths, which significantly exceed those of the optical Balmer lines. This motivated the exclusion of Ly$\alpha$ from the diagnostics in Figures \ref{fig:line_ratios}. We note that the observed dustiness of our obscured quasar sample may reveal it is biased towards objects with $P_{\rm{rad}}>>P_{\rm{hot}}$, since the impact of radiation pressure is increased by the presence of dust.

The additional dust observed in the Balmer line ratios of J1255$+$0339 could come from a foreground dust ``screen'' in the ISM of the quasar host, or could be due to a higher dust-to-gas ratio in the line-emitting clouds than our models assume. Additional extinction could move J1255$-$0339 towards the radiation pressure-dominated regime due to increased \ion{O}{6}/[\ion{O}{3}], since UV emission is subject to more attenuation than optical. It is possible the lack of UV line detection for this object is due to stronger dust extinction. This could be explained if J1255$+$0339 is in an early stage of AGN evolution characterized by high obscuration or even post-feedback quiescence \citep[``sunny weather"; see][for a review]{Gaspari_2020}. Higher-sensitivity optical measurements of lines such as \ion{Ne}{5} and \ion{Ne}{3} may help further characterizations of this object. 

\section{Conclusions}

This survey obtained UV, circum-nuclear spectra for five radio-quiet, low-$z$, luminous quasars exhibiting galactic-scale outflows in order to diagnose the mechanisms driving AGN feedback. By comparing \textit{HST} COS observations of narrow-line region, circum-nuclear emission lines in the FUV range to estimates of $P_{\rm rad}/P_{\rm hot}$, the ratio of hot gas pressure to radiation pressure predicted by hydrostatic photoionized AGN models, we constrain which mechanism currently dominates the dynamics of outflowing clouds.  

Our UV and optical diagnostics suggest that three of our five targets (J1356$+$1026, J0841$+$0101, and J1000$+$1242) fall within the range predicted by radiation pressure confinement models, reflecting the results of a pilot study by \cite{Somalwar:2020aa}. Another object's (J1222$+$0007) line ratio diagnostics may indicate the presence of a dynamically important hot wind, but fall relatively close to the $P_{\rm rad}/P_{\rm hot}=0$ model predictions, suggesting the possibility of non-negligible pressure contributions from both radiation and a hot wind, although the primary mechanism remains ambiguous with our data. The remaining target (J1255$+$0339) may include a dynamically important hot wind pressure contribution, but non-detections of \ion{O}{6} and \ion{C}{4} prevent a robust conclusion in this case. Our line ratios suggest the majority of our target outflows are driven by radiation pressure or were previously driven by a hot wind that is no longer dynamically important, though both theorized mechanisms could play a role in observed AGN feedback in our sample. \textcolor{black}{Constraining the contribution of a hot wind component and making broader conclusions about drivers of feedback from radio-quiet quasars more generally requires a larger sample of targets across a broad range of properties and morphologies observed in the UV and optical. Diagnosing dominant AGN feedback mechanisms across different stages of AGN evolution to make broad, population-level conclusions will likely require a multi-wavelength approach, particularly with mid-IR observations that are less susceptible to dust obscuration \citep{Ikiz_2020, Almeida_2025}.}

\section*{Acknowledgments}
This research is based on observations from the NASA/ESA Hubble Space Telescope obtained from the Space Telescope Science Institute, under NASA contract NAS 5–26555. 
Some of the data presented in this paper was obtained from the Mikulski Archive for Space Telescopes (MAST) at the Space Telescope Science Institute, dataset \dataset[https://doi.org/10.17909/27a2-6594]{https://doi.org/10.17909/27a2-6594}. These results are also based on observations from the European Southern Observatory (ESO) programmes 0103.B-0071 and 0104.B-0476 with DOIs https://doi.org/10.18727/archive/41 and https://doi.org/10.18727/archive/42.

This work is associated with program HST-GO-15935 and funded by grant number HST-GO-15935.021. STScI is operated by the Association of Universities for Research in Astronomy, Inc., under NASA contract NAS5–26555. Support to MAST for these data is provided by the NASA Office of Space Science via grant NAG5–7584. Basic research in radio astronomy at the U.S. Naval Research Laboratory is supported by 6.1 Base Funding. JIL is supported by the Eric and Wendy Schmidt AI in Science Postdoctoral Fellowship, a Schmidt Futures program. CAFG was supported by NSF through grants AST-2108230 and AST-2307327; by NASA through grants 21-ATP21-0036 and 23-ATP23-0008; and by STScI through grant JWST-AR-03252.001-A. M.G. acknowledges support from the ERC Consolidator Grant \textit{BlackHoleWeather} (101086804). EC is supported by the National Research Foundation of Korea (NRF-RS-2025-00515276).

\bibliographystyle{aasjournal}
\bibliography{main}



\end{document}